# Joint timing and frequency synchronization based on weighted CAZAC sequences for reduced-guard-interval CO-OFDM systems

Oluyemi Omomukuyo,[1] Deyuan Chang,[1] Jingwen Zhu,[1] Octavia Dobre,[1] Ramachandran Venkatesan,[1] Telex Ngatched,[2] and Chuck Rumbolt

[1]Faculty of Engineering and Applied Science, Memorial University, St. John's, NL, A1B 3X5, Canada
[2]Division of Science, Grenfell Campus, Memorial University, Corner Brook, NL, A2H 5G4, Canada

**Abstract:** A novel joint symbol timing and carrier frequency offset (CFO) estimation algorithm is proposed for reduced-guard-interval coherent optical orthogonal frequency-division multiplexing (RGI-CO-OFDM) systems. The proposed algorithm is based on a constant amplitude zero autocorrelation (CAZAC) sequence weighted by a pseudo-random noise (PN) sequence. The symbol timing is accomplished by using only one training symbol of two identical halves, with the weighting applied to the second half. The special structure of the training symbol is also utilized in estimating the CFO. The performance of the proposed algorithm is demonstrated by means of numerical simulations in a 115.8-Gb/s 16-QAM RGI-CO-OFDM system.

## 1. Introduction

In recent years, orthogonal frequency division multiplexing (OFDM) has become an attractive technology for high-speed optical communication systems because it offers high spectral efficiency and high tolerance to both the fiber chromatic dispersion (CD) and polarization mode dispersion (PMD) [1–3]. Coherent optical OFDM (CO-OFDM) has demonstrated superior transmission performance in terms of spectral efficiency and receiver sensitivity than its direct-detection counterpart [1], making it more suitable for long-haul transmission. Conventional CO-OFDM systems utilize a cyclic prefix (CP) between adjacent OFDM symbols to accommodate the inter-symbol interference (ISI) arising from the fiber CD and PMD [4,5]. Given that CD increases quadratically with an increase in the data rate [6], the large CD-induced channel memory length poses a problem for long-haul, high-speed transmission, especially when there is no in-line optical dispersion compensation [7]. This is because a long CP duration would be required to compensate for the CD, resulting in a larger overhead and poorer spectral efficiency. To overcome this problem, reduced-guard-interval CO-OFDM (RGI-CO-OFDM) [7] has been proposed. RGI-CO-OFDM systems employ a reduced guard interval to accommodate only the PMD, while the CD is compensated using frequency-domain equalization (FDE) at the receiver similar to single-carrier (SC) systems.

Despite its several advantages, OFDM systems suffer from a higher degree of sensitivity to frequency synchronization errors when compared with SC systems [8]. OFDM systems require timing and frequency synchronization before an accurate symbol decision can be made at the receiver. Timing synchronization entails finding an estimate of the correct position of the Discrete Fourier Transform (DFT) window at the receiver so as to avoid ISI, while frequency synchronization involves estimating and compensating for the carrier frequency offset (CFO) which may cause inter-carrier interference (ICI) between the OFDM subcarriers. In wireless communications, several algorithms for carrying out OFDM timing and frequency synchronization either jointly or individually have been proposed [9-12]. These algorithms can be classified into data-aided (DA) [9-11] and non-data-aided (NDA) [12] methods. In DA algorithms, which are the focus of this paper, the timing and frequency synchronization is usually based on exploiting the correlation property of specially-designed training symbols (typically with some sort of repetitive pattern).

One of the most popular DA algorithms, proposed by Schmidl and Cox [9], uses a training symbol with two identical halves for the timing synchronization. However, the timing metric of the Schmidl and Cox's algorithm has a plateau which results in a large timing offset estimation variance. In order to eliminate the timing metric plateau of the Schmidl and Cox's algorithm, Minn *et al*. [10] proposed a modified training symbol with four identical parts having specific sign changes for these parts. The resulting timing metric has a steeper rolloff, but has a large timing estimation variance in ISI channels. Park *et al*. [11] proposed an algorithm based on reverse autocorrelation, which uses a repeated-conjugated-symmetric sequence to improve the timing estimation variance of the Minn's algorithm. Although the Park's algorithm results in an impulse-shaped timing metric which yields a more accurate timing offset estimation, the timing metric has two large side lobes which can result in errors in the timing synchronization. Some of these algorithms have been considered for timing synchronization in coherent optical systems [13-16].

Unlike in conventional OFDM wireless systems, where the CFO is usually due to the Doppler effect, the CFO in CO-OFDM systems is brought about by the incoherence of the signal laser of the transmitter and the local oscillator (LO) laser of the receiver. Since commercially-available lasers are usually locked to an International Telecommunication Union (ITU) standard, but only with a frequency accuracy within ±2.5 GHz over their lifetime [17], the CFOs in CO-OFDM systems is typically within the range [-5 GHz, +5 GHz]. Given that optical phase-locked loops are disadvantaged by high cost and complexity, frequency synchronization algorithms are essential to CO-OFDM receivers. The Schmidl and Cox's algorithm can carry out frequency synchronization by computing the phase difference between the two halves of the training symbol. However, the CFO estimation range is limited to ±subcarrier spacing, making it unsuitable for use in CO-OFDM systems. The CFO estimation range can however be increased by employing a second training symbol, but at the cost of extra overhead [9].

In this paper, we propose and demonstrate, for the first time to our knowledge, a joint timing and frequency synchronization algorithm for RGI-CO-OFDM systems using only one training symbol based on a constant amplitude zero autocorrelation (CAZAC) sequence weighted by a pseudo-random noise (PN) sequence. CAZAC sequences, which are a type of polyphase codes, have constant amplitude elements and good periodic autocorrelation properties [18-20] and have been applied in wireless communications systems for various applications, including channel estimation [21] and synchronization [22-24]. The performance of the proposed technique is demonstrated by means of numerical simulations in a 115.8-Gb/s 16-QAM RGI-CO-OFDM system. The proposed algorithm is shown to have a wide CFO estimation range, as well as a more precise timing offset estimation and a better CFO estimation performance than popular existing synchronization methods.

## 2. Principle of proposed synchronization algorithm

The proposed synchronization algorithm makes use of a training symbol of two identical halves, with the second half weighted by a PN sequence. Each half of the training symbol has a length of $M = N/2$, and is generated by an $M$-point inverse fast Fourier transform (IFFT) of a CAZAC sequence of length $L = N_{sc}/2$, where $N_{sc} (\leq N)$ is the number of OFDM subcarriers, and $N$ is the IFFT size.

The CAZAC sequence, $c(m)$ can be expressed as:

$$c(m) = \exp\left(\frac{j\pi r m^2}{L}\right) \qquad m = 0, 1, \cdots, L-1, \qquad (1)$$

where $r$ is a positive integer which is relative-prime to $L$ [19]. The autocorrelation property of the CAZAC sequence is:

$$\sum_{m=0}^{L-1} c(m) c^*\left([m+\tau]_{\bmod L}\right) = \begin{cases} L, & \tau = 0 \\ 0, & \tau \neq 0 \end{cases} \qquad (2)$$

where the superscript * represents the complex conjugation operation and $[\square]_{\bmod L}$ is the modulo-$L$ operator. The structure of the proposed training symbol is:

$$TS = [A_M \ B_M], \quad (3)$$

where $A_M$ represents the $M$-point IFFT of $c(m)$ and $B_M$ is obtained by multiplying $A_M$ by a real-valued PN sequence, $p(n) \in (-1,1]$, where $n = 0, 1, \cdots, M-1$. Note that $p(n)$ is introduced to scramble the samples in the second half of the training symbol so as to eliminate the timing metric plateau associated with the Schmidl and Cox's algorithm. In this regard, a PN sequence of all "1s" or all "-1s" will not work in eliminating this plateau. The samples in the second half of the training symbol have to be later descrambled in the receiver.

*2.1 Timing synchronization*

The timing synchronization is based on the timing metric, $M(d)$, which is defined as:

$$M(d) = \frac{|P(d)|^2}{R^2(d)}, \quad (4)$$

with

$$P(d) = \sum_{n=0}^{M-1} r(d+n) p(n) r^*(d+n+M), \quad (5)$$

and

$$R(d) = \frac{1}{2} \sum_{k=0}^{N-1} |r(d+k)|^2, \quad (6)$$

where $r(n)$ represents the discrete samples of the received OFDM signal, $d$ is the time index corresponding to the first received sample in a window of $N$ samples, $P(d)$ represents the cross-correlation between the two halves of the window, and $R(d)$ is the half-symbol energy in the $N$ samples of the window. The timing metric defined in Eq. (4) contains two modifications to the timing metric of the Schmidl and Cox's algorithm. The first modification is the introduction of $p(n)$ in Eq. (5) to descramble the samples in the second half of the training symbol. For the second modification, all samples over one symbol period are used in computing $R(d)$ instead of using only the samples in the second half symbol period. The timing offset estimate is obtained as the time index at which $M(d)$ has its peak value:

$$\hat{d} = \arg[\max_d (M(d))]. \quad (7)$$

*2.2 Frequency synchronization*

After transmission through the optical fiber, if the laser phase noise and amplified spontaneous emission (ASE) noise are both neglected, and it is assumed that the phase shift induced by the CD has been compensated for, the two halves of the training symbol will differ by $p(n)$ and a phase shift induced by the CFO.

The CFO $\Delta f$ can be decomposed into a fractional part with a magnitude $\leq \Delta f_N$, and an integer part, which is a multiple of $2\Delta f_N$, where $\Delta f_N$ is the OFDM subcarrier frequency spacing [9]. If we let $\rho = \frac{\Delta f}{\Delta f_N}$ be the normalized CFO, $\rho$ can be expressed as:

$$\rho = \alpha + 2\beta, \tag{8}$$

where $\alpha$ is the normalized fractional CFO and $2\beta$ is the normalized integer CFO, with $|\alpha| \leq 1$ and integer $\beta$ spanning the range of possible CFOs. It is straightforward to express the received samples in the second half of the training symbol as:

$$r(n+M) = p(n)r(n)e^{j\pi\alpha}. \tag{9}$$

Equation (9) shows that the two halves of the training symbol differ only by the PN sequence and a phase shift of $\pi\alpha$. Consequently, if timing synchronization has already been carried out, an estimate, $\hat{\alpha}$, of $\alpha$ can be obtained as:

$$\hat{\alpha} = \frac{-1}{\pi} angle\left(\sum_{n=0}^{M-1} r(\hat{d}+n)p(n)r^*(\hat{d}+n+M)\right). \tag{10}$$

Equation (10) indicates that the correct timing information is necessary to compute $\hat{\alpha}$. In order to estimate the integer CFO, the samples of the training symbol have to be first counter-rotated at an angular speed of $2\pi\hat{\alpha}\Delta f_N t$, where $0 \leq t \leq T$ and $T$ is the useful OFDM symbol duration. By compensating for the fractional CFO, ICI is eliminated, and there is no loss of orthogonality among the OFDM subcarriers. However, because of the uncompensated integer CFO, the fast Fourier transform (FFT) outputs will be shifted by $2\beta$. In order to obtain $\beta$, we compute the correlation in the frequency domain of the fractional CFO-compensated training symbol with the original transmitted training symbol.

Let the FFT of the received training symbol with only integer CFO be $R_f(k)$ and let the FFT of the original training symbol be $B_f(k)$. We can define the normalized cross-correlation between $R_f(k)$ and $B_f(k)$ as:

$$\Psi(\beta) = \frac{\left|\sum_{k=0}^{N-1} B_f^*(k) R_f(k+2\beta)\right|^2}{\left(\sum_{k=0}^{N-1} |B_f(k)|^2\right)^2} \quad \beta = -\frac{M}{2}, -\frac{M}{2}+1, \cdots, \frac{M}{2}-1. \tag{11}$$

The estimate, $\hat{\beta}$, of $\beta$ is obtained as the index that maximizes $\Psi(\beta)$:

$$\hat{\beta} = \arg\left[\max_\beta (\Psi(\beta))\right]. \tag{12}$$

The estimate of the combined normalized CFO would then be:

$$\hat{\rho} = \hat{\alpha} + 2\hat{\beta}. \tag{13}$$

From the above, it can be deduced that the CFO estimation range of the proposed algorithm is $(-M \leq \hat{\rho} \leq M-1)\Delta f_N$.

## 3. Simulation setup

The simulation schematic of the RGI-CO-OFDM system used to investigate the performance of the proposed synchronization algorithm is shown in Fig. 1(a). The digital signal processing (DSP) at the transmitter and receiver is performed in MATLAB while the optical system model is built using VPI TransmissionMaker. At the transmitter, a $2^{19}$ deBruijn sequence is generated and then mapped onto the OFDM subcarriers with 16-ary quadrature amplitude modulation (16-QAM). The time-domain RGI-CO-OFDM signal is generated using a 512-point IFFT with a CP length of 9% to accommodate the ISI induced by the fiber PMD and to increase the tolerance of the system to synchronization errors. Of the 512 channels, 412 are data subcarriers, 99 are zero-valued edge subcarriers for ~ 20% oversampling to combat aliasing, and one zero-valued subcarrier is reserved for the DC term. After CP insertion, the training symbol used for synchronization is inserted at the beginning of the OFDM frame, and one training symbol is employed every 50 data symbols for channel estimation, resulting in a training symbol overhead of ~ 2% [25]. For the synchronization training symbol, $r$ is chosen to be $L-1$. The structure of the OFDM frame is shown in Fig. 1(b).

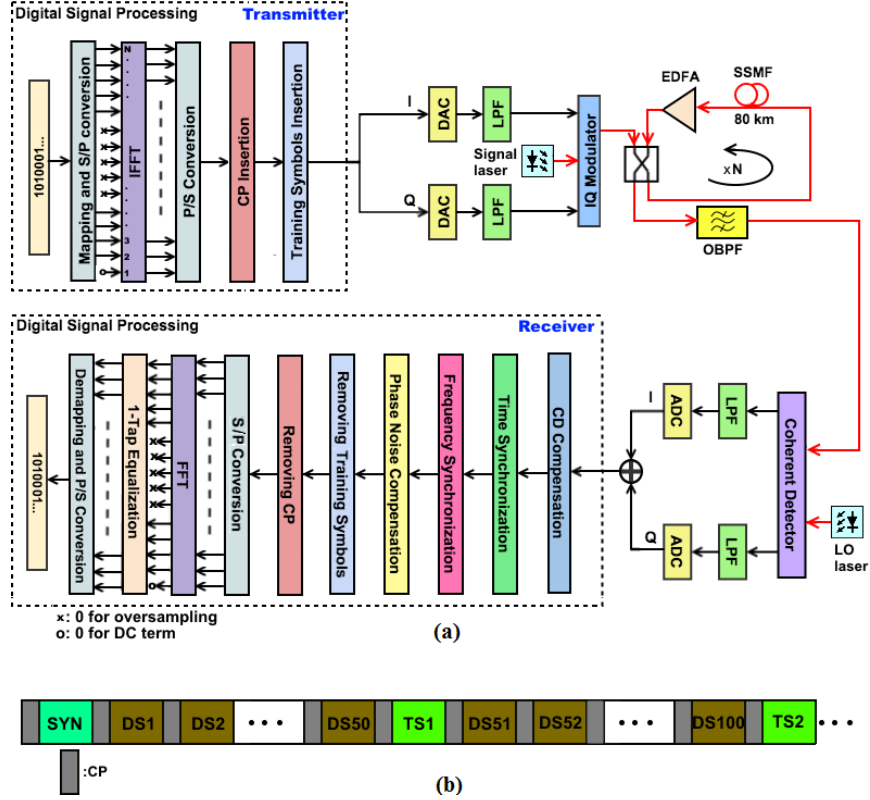

Fig. 1. (a) Simulation setup of the RGI-CO-OFDM system. (b) OFDM frame structure. S/P: serial-to-parallel conversion. P/S: parallel-to-serial conversion. LPF: low-pass filter. OBPF: optical band-pass filter. SYN: synchronization symbol. DS: data symbol. TS: training symbol.

Although the schematic depicted in Fig. 1 is for a single polarization system, it can be extended for dual-polarization transmission. In such a case, an identical CAZAC training symbol would be inserted at the beginning of each OFDM frame for each polarization. The proposed algorithm would then be used for the synchronization of the OFDM frames in the two polarization branches as described in Section 2.

Taking into account the overheads from the CP and the training symbols, the net data rate of the RGI-CO-OFDM signal is 115.8 Gb/s {40 GSa/s×4×(412/512)×[1/(1.02×1.09)]} and the OFDM subcarrier spacing is 78.125 MHz (40 GHz/512). The real and imaginary parts of the RGI-CO-OFDM signal are loaded to digital-to-analog converters (DACs) operating at 40 GSa/s and then used to drive an I/Q modulator whose sub Mach-Zehnder modulators (MZMs) are biased at the transmission null. The I/Q modulator is used to modulate the signal laser to generate the optical RGI-CO-OFDM signal. The signal laser is a continuous wave (CW) laser with a linewidth of 100 kHz, center emission wavelength of 1550 nm and average output power of 0 dBm.

The optical signal is launched into a recirculating loop consisting of 80-km standard single mode fiber (SSMF) and an erbium-doped fiber amplifier (EDFA). The gain and noise figure of the EDFA are 16 dB and 4 dB, respectively. The CD parameter, PMD coefficient, loss and nonlinearity of the fiber are 16 ps/nm/km, 0.1 ps/√km, 0.2 dB/km and $2.6 \times 10^{-20}$ $m^2$/W, respectively. An optical band-pass filter (OBPF) with bandwidth of 0.8 nm is used for ASE suppression.

At the receiver, the RGI-CO-OFDM is mixed with the LO laser operating in CW mode with a linewidth of 100 kHz, and then detected by a coherent receiver comprising a 2×4 quadrature optical hybrid and two pairs of balanced photodiodes. Next, the coherently-detected signal is digitized by analog-to-digital converters (ADCs) at 40 GSa/s with 8-bit resolution prior to CD compensation. A frequency-domain equalizer using the overlap-add method [26] is utilized for CD compensation. Time and frequency synchronization is accomplished by the proposed algorithm while the RF-pilot method [25] is used for phase noise compensation. The training symbols are utilized for channel estimation and a one-tap equalizer is employed after the FFT to compensate for the PMD and any residual CD. The OFDM symbols are then demapped and the bit error rate (BER) is obtained by direct error counting.

## 4. Simulation results

*4.1 Timing synchronization performance*

Figure 2(a) shows the measured timing metric of the proposed algorithm for 800-km SSMF transmission (with CD compensation), without optical noise. There is also no CFO between the signal and LO lasers. At the receiver, a timing offset is modeled as a delay in the received signal. The correct timing instant, indexed 0 in the figure, is the start of the useful part of the training symbol. The timing metrics corresponding to the Schmidl and Cox's and the Minn's algorithms are also included for comparison. As can be seen in Fig. 2(a), the timing metric of the Schmidl and Cox's method maintains a plateau for the entire CP length (46 samples). It is evident that this plateau results in some uncertainty as to the actual start of the DFT window. Unlike the Schmidl and Cox's method, the timing metric of the Minn's method has a triangular shape with no trajectory plateau. The timing metric obtained using the proposed algorithm is impulse-shaped with no sidelobes, and has a sharp peak at the correct timing instant while the values are almost zero at all other positions.

Figure 2(b) shows the measured timing metrics of the three estimators in the absence of optical noise but with a CFO of 5 GHz. It can be seen that unlike the timing metrics of the other two estimators, in the presence of such a large CFO, the peak value of the timing metric of the proposed algorithm has reduced from its ideal value of 1. However, the timing metric of the proposed method still maintains its impulse shape, with the peak at the correct timing instant, implying that no timing uncertainty is brought about by the CFO.

Figure 2 (c) shows the timing metrics of the three estimators when the optical signal-to-noise ratio (OSNR) and CFO are 6 dB and 5 GHz, respectively. It can be seen that the timing metric of the Schmidl and Cox's method has deteriorated significantly in the presence of the high level of optical noise. This would result in a large timing uncertainty. The timing metric of the Minn's method is also affected by optical noise, but to a lesser degree than the Schmidl and Cox's. Although the Minn's timing metric still has a triangular shape, there is now a shift at the top of the triangle which would result in timing uncertainty. In contrast, the timing metric of the proposed method is still impulse-shaped, allowing it to achieve a more accurate timing offset estimation even at low OSNR levels.

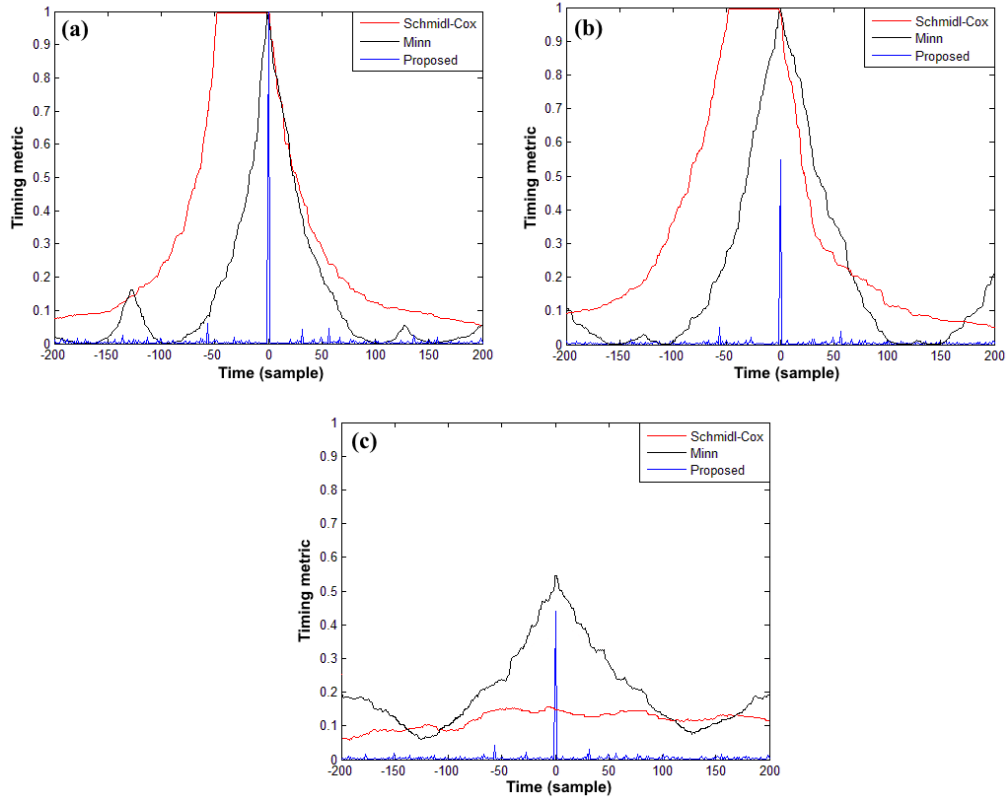

Fig. 2. Comparison of timing metric of estimators for 800-km SSMF transmission. (a) without CFO and without optical noise. (b) with a CFO of 5 GHz and without optical noise. (c) with a CFO of 5 GHz and for an OSNR of 6 dB.

Figures 3 and 4 show the means and variances of the timing estimators versus OSNR for 800-km SSMF transmission and with a CFO of 5 GHz. It can be seen from Fig. 3 that even at high levels of OSNR, the mean value of the Schmidl and Cox's method is within the CP, yielding a high timing estimation variance as illustrated in Fig. 4. Further reduction in the OSNR results in more errors in the mean timing offset estimation and consequently, a larger timing estimation variance. The Minn's algorithm keeps the correct timing estimation with small timing estimation variance at high OSNR levels, but starts to yield inaccuracies at OSNR values less than 14 dB. The proposed method gives a more accurate timing estimation than the other methods over the range of considered OSNRs. In addition, since no timing offset variations are observed for the proposed method, the timing estimation variance is not included in the results of Fig. 4.

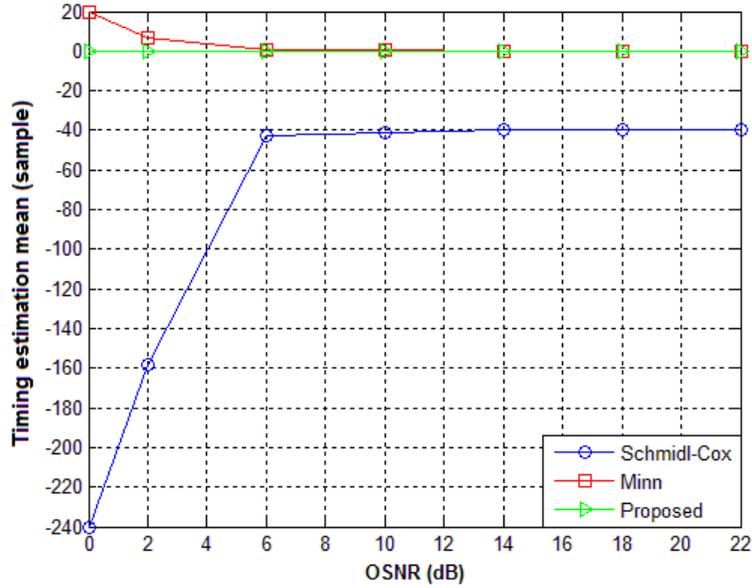

Fig. 3. Timing estimation mean vs. OSNR for 800-km SSMF transmission with a CFO of 5 GHz.

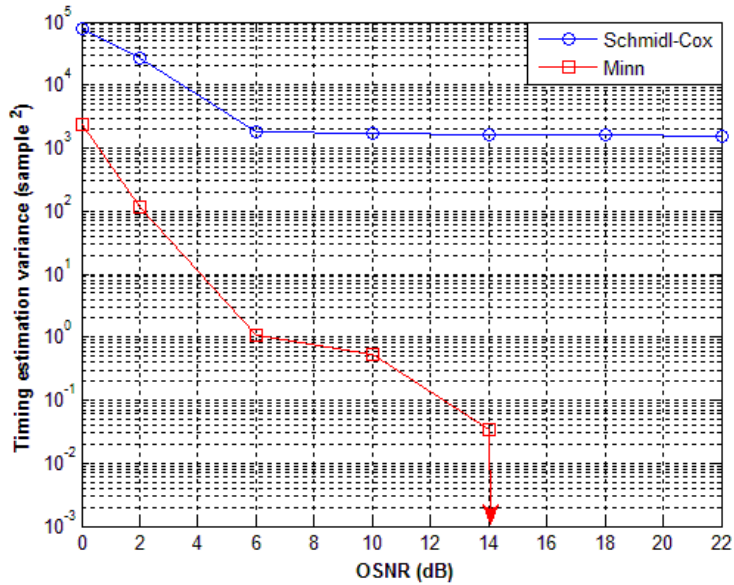

Fig. 4. Timing estimation variance vs. OSNR for 800-km SSMF transmission with a CFO of 5 GHz (no timing offset variations are observed for the proposed method, hence, the corresponding results are not included in the figure).

*4.2 Frequency synchronization performance*

For the investigations on the frequency synchronization performance of the proposed method, the maximum CFO between the signal and LO lasers is limited to $\pm 5$ GHz and the SSMF length is fixed at 800 km. In addition, the investigations are carried out in the presence of the original timing offset used in the simulations in Section 4.1. Figure 5 shows the mean of the estimated CFO as a function of the actual CFO for an OSNR of 18 dB. For comparison, we have also included the CFO estimation performances of the Schmidl and Cox's algorithm, as well as the RF-pilot aided frequency offset estimation (RAFOE) scheme proposed by Zhou *et al*. [13].

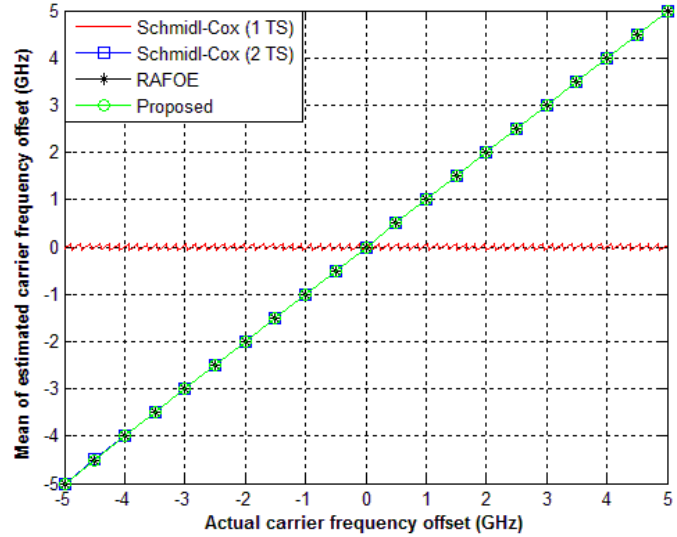

Fig. 5. Mean of estimated CFO vs. actual CFO for 800-km SSMF transmission and an OSNR of 18 dB.

Figure 6 shows a zoomed-in version of Fig. 5. The Schmidl and Cox's algorithm utilizes the same training symbol employed for timing synchronization and for fractional CFO estimation.

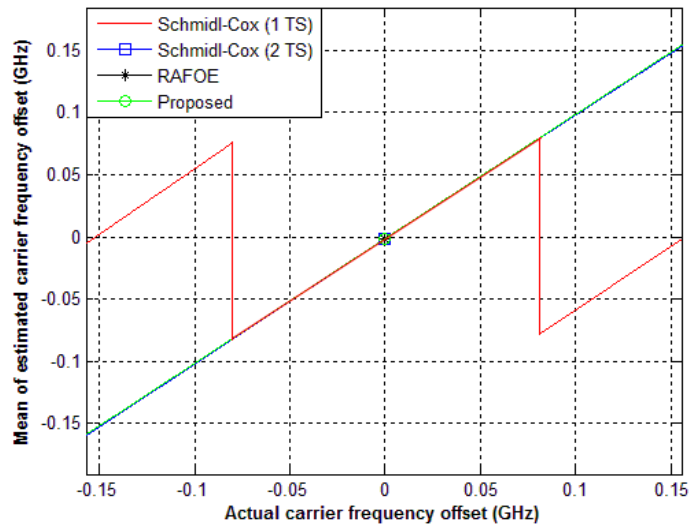

Fig. 6. Zoomed-in version of Fig. 5, illustrating the CFO estimation range of the Schmidl and Cox's algorithm when 1 TS is used.

It can be observed from Fig. 6 that the CFO estimation range of the Schmidl and Cox's algorithm when this training symbol is used is limited to ±78.125 MHz. This would render it unsuitable for CFO estimation in high-speed CO-OFDM systems, unless a second training symbol is employed to measure integer CFOs and thus increase the CFO estimation range, as seen in Fig. 5. In contrast, the CFO estimation range of the proposed method is -20 GHz ≤ $\rho$ ≤ 19.92 GHz, allowing it to comfortably estimate CFOs within the maximum expected range of ±5 GHz. The RAFOE algorithm also uses the first training symbol of the Schmidl and Cox's method both for timing synchronization and for estimating the fractional CFO, while the integer CFO is estimated using an RF-pilot. The theoretical estimation range of the RAFOE algorithm is as wide as half of the sampling rate [13]. This also makes it suitable for estimating CFOs in CO-OFDM systems without requiring any extra training symbol overhead.

In order to illustrate the accuracy of the CFO estimation of the three methods, the mean square error (MSE) of the CFO estimation in the presence of optical noise has been obtained as shown in Fig. 7 for a CFO of 5 GHz. The second training symbol has been employed for the Schmidl and Cox's algorithm to increase its CFO estimation range to cover 5 GHz. It is clear from Fig. 7 that the proposed method has a smaller MSE and hence a more accurate CFO estimation than the other algorithms. The improvement in the MSE performance shown by the proposed algorithm can be attributed to the more accurate timing offset estimation it demonstrates, and the ample phase information contained in the two halves of the training symbol. A combination of these two factors would result in a more accurate fractional CFO estimation and consequently, a more accurate combined CFO estimate. Since the RAFOE algorithm uses the first training symbol of the Schmidl and Cox's algorithm for joint timing synchronization and for fractional CFO estimation, it shows a similar MSE performance as the Schmidl and Cox's method.

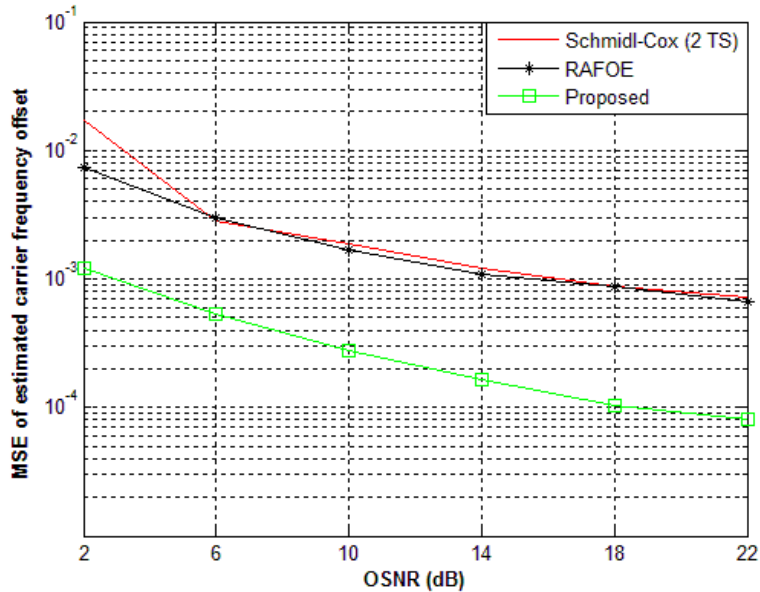

Fig. 7. MSE of the estimated CFO vs. OSNR for 800-km SSMF transmission and a CFO of 5 GHz.

Figure 8 shows the BER (post-CFO compensation) as a function of the given CFO using the proposed algorithm for OSNRs of 18 dB and 22 dB. It is important to state that the BER values are obtained without implementing forward error correction (FEC). The results of Fig.

8 show that for the different levels of OSNR, the BER basically remains constant over the range of CFOs considered.

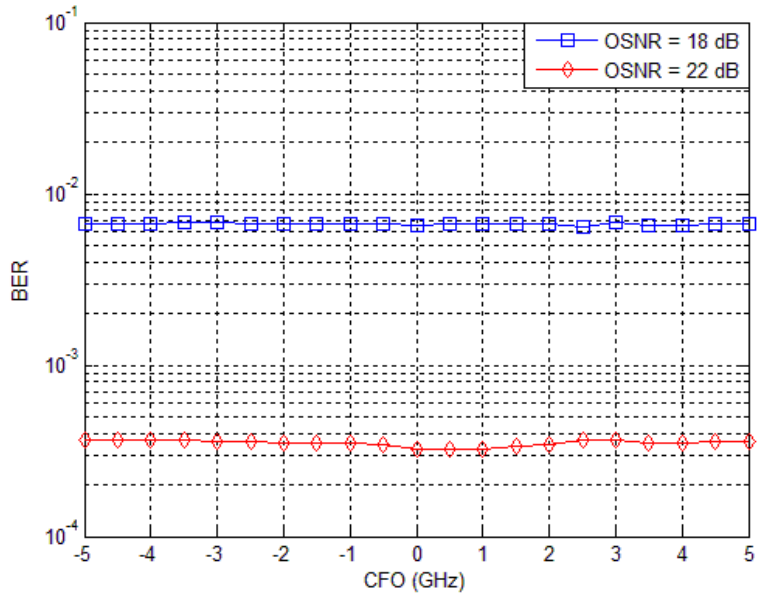

Fig. 8. BER vs. CFO for 800-km SSMF transmission.

Figure 9 shows the BER against the OSNR for the cases when there is no CFO, and when there is a CFO of 5 GHz compensated for by using the three methods. It can be seen that for all the OSNR values, the proposed method has virtually the same BER performance as the system with no CFO. In contrast, there is a noticeable OSNR penalty for the other methods.

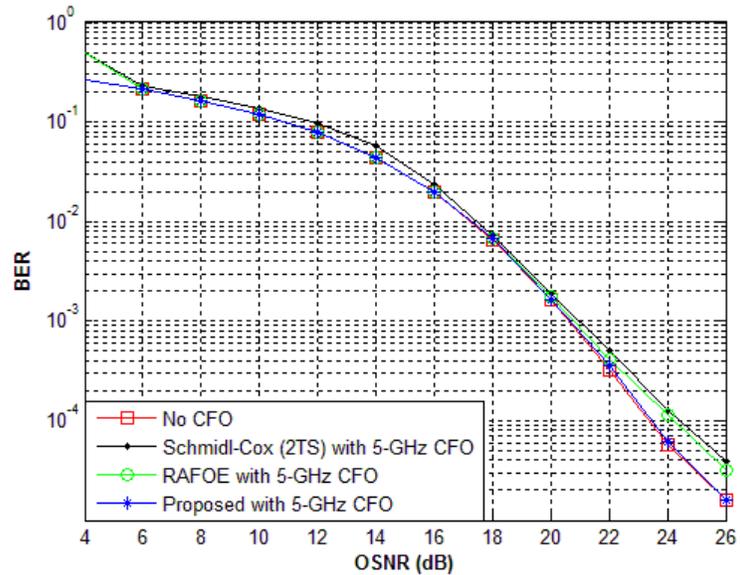

Fig. 9. BER vs. OSNR for 800-km SSMF transmission.

## 5. Conclusion

A novel joint timing and frequency synchronization algorithm using only one training symbol based on a weighted CAZAC sequence has been proposed, and its performance numerically investigated in a 115.8-Gb/s 16-QAM RGI-CO-OFDM system. The proposed algorithm has a wide CFO estimation range and has demonstrated better timing and CFO estimation performance even at low OSNR values when compared with popular existing synchronization algorithms.

**Acknowledgment**

This work has been supported by the Atlantic Canada Opportunities Agency (ACOA), Research and Development Corporation (RDC), and Altera Corporation.